\documentclass[preprints,article,accept,moreauthors,pdftex10pt,a4paper]{Definitions/mdpi} 

\firstpage{1} 
\pubvolume{xx}
\issuenum{1}
\articlenumber{5}
\pubyear{2019}
\copyrightyear{2019}
\history{Received: 19 December 2018; Accepted: 5 February 2019; Published: 11 February 2019}

\Title{Magnetic field vector maps of nearby spiral galaxies}

\Author{Hiroyuki Nakanishi $^{1}$, Kohei Kurahara $^{1}$, and Kenta Anraku$^{1}$}
\AuthorNames{H. Nakanishi, K. Kurahara, and K. Anraku}

\address{%
$^{1}$ \quad Graduate School of Science and Engineering, Kagoshima university, 1-21-35 Korimoto, Kagoshima 890-0065, Japan; hnakanis@sci.kagoshima-u.ac.jp}

\corres{Correspondence: hnakanis@sci.kagoshima-u.ac.jp; Tel.: +81-99-285-8963}

\abstract{We present a method for determining directions of magnetic field vectors in a spiral galaxy using two synchrotron polarization maps, an optical image, and a velocity field. The orientation of the transverse magnetic field is determined with a synchrotron polarization map of higher frequency band and the $180^\circ$-ambiguity is solved by using sign of the Rotation Measure (RM) after determining geometrical orientation of a disk based on a assumption of trailing spiral arms. The advantage of this method is that direction of magnetic vector for each line of sight through the galaxy can be inexpensively determined with easily available data and with simple assumptions.
We applied this method to three nearby spiral galaxies using archival data obtained with the Very Large Array (VLA) to demonstrate how it works. 
The three galaxies have both clockwise and counter-clockwise magnetic fields, which implies that all three galaxies are not classified in simple Axis-Symmetric type but types of higher modes and that magnetic reversals commonly exist.}

\keyword{ISM: magnetic fields; galaxies: spiral; galaxies: magnetic fields; radio continuum: ISM; methods: data analysis}

\begin{document}
\section{Introduction}
A large-scale magnetic field of 1--10 $\mu$G exists in a spiral galaxy and is generally aligned with spiral arms \citep{sof86}. 
Cosmic ray electrons produce synchrotron emission by spiraling around such ordered magnetic field. Consequently, synchrotron emission is polarized transversely to magnetic field. Polarization observation thus provides us with an information on magnetic orientation. However, there is the $180^\circ$-ambiguity in the direction of magnetic field as far as an observation is carried out in a single frequency band. The line-of-sight component of the magnetic field and thermal electrons make polarization angle rotate, which is called Faraday rotation. The rotational angle is proportional to the square of wavelength and its slope is called Rotation Measure (RM), which is proportional to the integrated strength of magnetic field and thermal electron density along the path of line-of-sight. The positive RM indicate that the magnetic field is directed to the observer and the negative RM indicates the opposite direction. The RM measured with multi-frequency radio polarization observations tells us the line-of-sight direction and strength of the magnetic field. 

The morphology of magnetic fields in galaxies can be classified based on its dominant mode: Axis-Symmetric Spiral (ASS, $m=0$), Bi-Symmetric Spiral (BSS, $m=1$), and Quadri-Symmetric Spiral (QSS, $m=2$) types \citep{aka18}.
The azimuthal changes of RM can be used to classify the morphology and to solve the $180^\circ$-ambiguity \citep{aka18}.
Former works have shown that IC342 and M31, for example, are classified into ASS \citep{gra88,sof87} and that M81, M33 and NGC 6946 are classified into BSS \citep{kra89,bec91} by examining azimuthal changes of RM. 
However, recent high-resolution observations for nearby galaxies shows that RM distribution cannot be explained with simple ASS, BSS, or QSS models \citep{soi02,fle11}. {\citet{ste08} present models for the ionized gas and magnetic field patterns to explore which modes exist in a magnetic field fitting observational data. }

Therefore, our knowledge on the magnetic field configurations is still incomplete though magnetic field configurations are fundamental information for galaxies and have possibilities of being related to its origin \citep{sof86,sof10}. 
In this paper, we propose a new method to investigate the structure of magnetic field based on simple assumptions. The first idea of this method appeared originally in a master's thesis written by \citet{anr14}. 
The following section 2 describes selections of objects to apply the method, data reduction, and magnetic field maps. 
The section 3 is devoted to details of our method for determining the direction of magnetic field vector. Section 4 comprised our results of vector maps of magnetic fields. 
In section 5 we discuss the configurations of magnetic fields and section 6 provides a summary. 

Note that we discriminate the meanings of ``orientation'' and ``direction'' in this paper; ``orientation'' is defined as an angle in the range of $0^\circ$ -- $180^\circ$ with $180^\circ$ ambiguity and ``direction'' is defined as an angle in the range of $0^\circ$ -- $360^\circ$ without any ambiguity.

\section{Data}
\subsection{Selection of Objects}
We selected galaxies from Nearby Galaxy Catalog (NBG) \citep{tul88} with the following criteria (1) heliocentric distance is within 4--10 Mpc to cover an entire disk with a single pointing and to map with a resolution of less than a few kpc, (2) inclination range of $30^\circ$ -- $60^\circ$, (3) brighter than the B magnitude of 11, and (4) observed in C and X bands with the VLA and listed in Table 5 of \citet{bec13}. 
Selected galaxies meeting these criteria (1)--(4) were NGC 4414, NGC 4736, and NGC 6946. Table \ref{tb:gal} lists basic parameters of these three galaxies; Name, Positions in the equatorial coordinates, Morphology, B-band magnitude, Recession velocity, inclination, and position angle.

\subsection{Data Reduction}
We used the software AIPS (Astronomical Image Processing System) developed by NRAO (National Radio Astronomy Observatory) for polarization data reduction \citep{gre98}. 

The basic procedure of the primary data reduction was as follows: (1) loading FITS files (AIPS task: FILLM), (2) flagging first 10-second data of each scan and spurious data (QUACK and TVFLG), (3) entering the source flux density into SU (source table) (SETJY), (4) determining antenna based calibration using the calibrator sources (VLAPROCS, VLACALIB), (5) determining source flux densities (GETJY), (6) applying calibration using SN (solution table) (VLACLCAL), (7) applying D-term calibration (PCAL), and finally (8) correcting the phase difference between right and left polarizations (RIDIF). Thus, the visibility data were calibrated in this primary reduction. 

After the primary data reduction, we made images of Stokes I, Q, and U by applying the task IMAGR to the visibility data. 
Since the obtained maps have different pointing centers and fields of view depending on the frequency, we aligned the coordinates of images using the task HGEOM. Finally, polarization angle $\psi$ and polarization intensity $P$ were calculated with $\psi=\frac{1}{2}\arctan\frac{U}{Q}$ and $P=\sqrt{Q^2+U^2}$, respectively, using the task COMB. 

\subsection{Maps of Magnetic Fields}

\begin{figure}[h]
\centering
\includegraphics[height=5 cm]{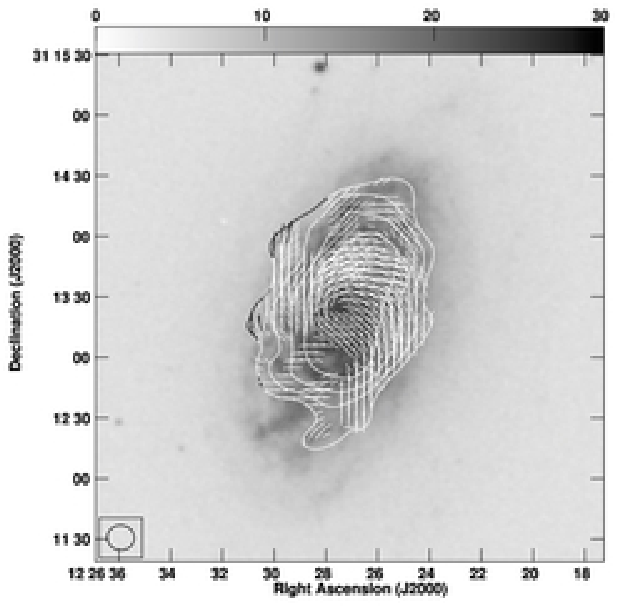}
\includegraphics[height=5 cm]{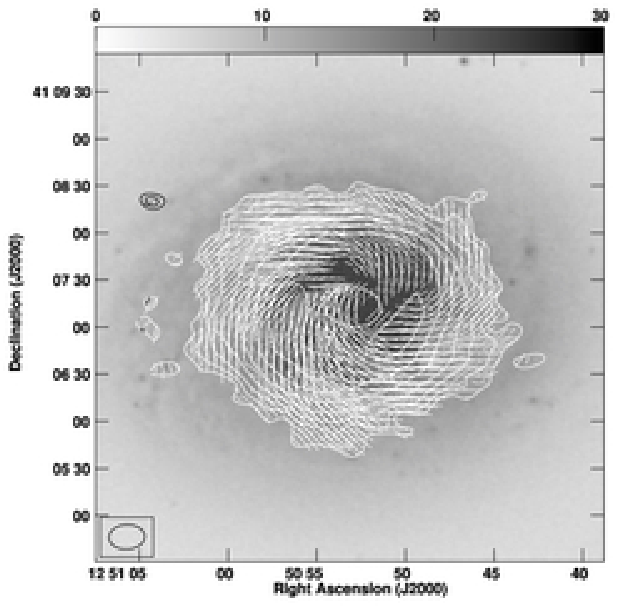}
\includegraphics[height=5 cm]{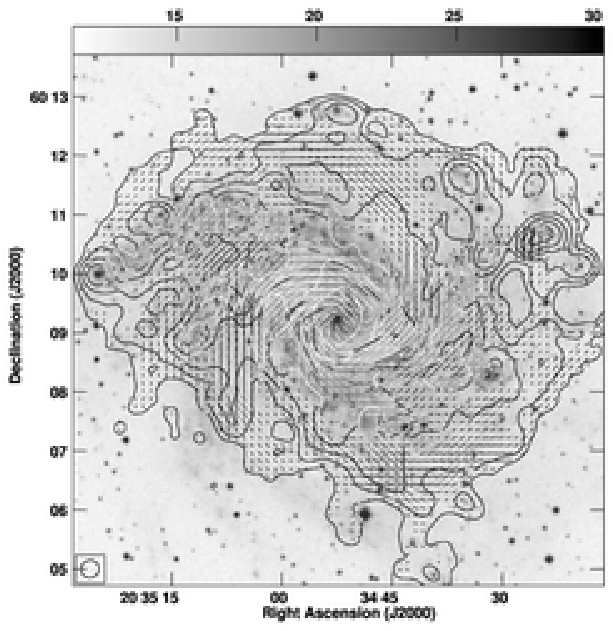}
\includegraphics[height=5 cm]{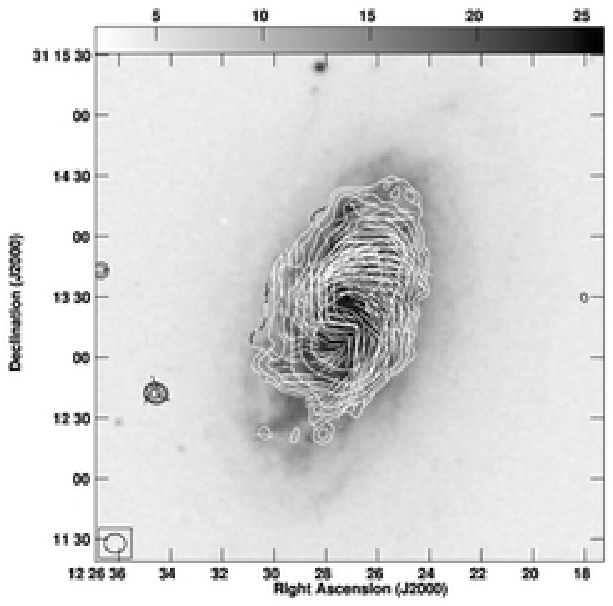}
\includegraphics[height=5 cm]{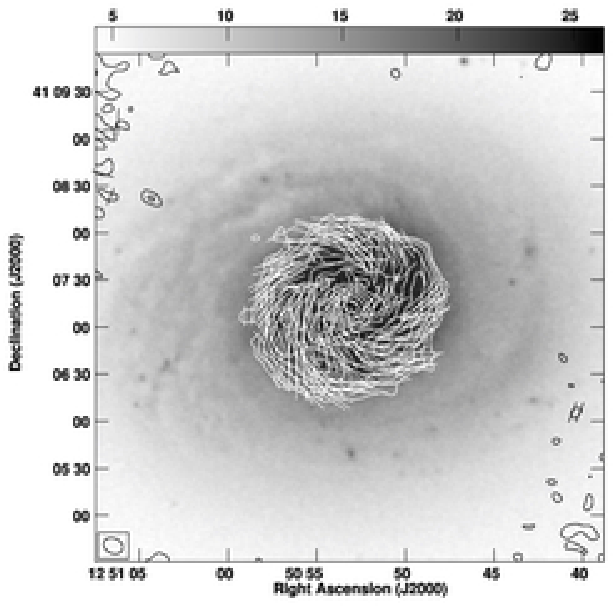}
\includegraphics[height=5 cm]{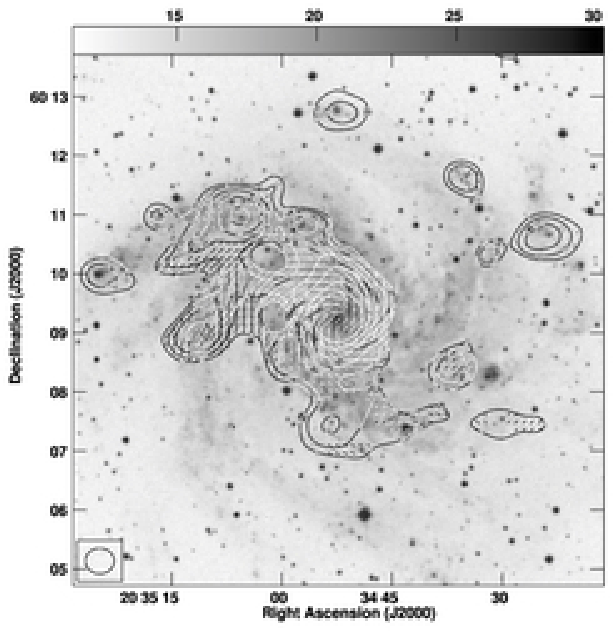}
\caption{Magnetic orientation maps superposed on contour plots of the Stokes I for NGC 4414 (Left), NGC 4736 (Middle), and NGC 6946 (Right) in the C (Top) and X (Bottom) bands. Grey scale and contour images are B-band optical images taken from the DSS (Digitized Sky Survey) and images of Stokes I. Contour levels are 3, 6, 12, 24, 48, and $96\times \Delta I$ (mJy/beam). Pixel numbers are 1024 for all the images and magnetic field orientations are plotted every 15 pixels. }
\label{mag-ori}
\end{figure}

Maps of the obtained magnetic fields for three galaxies are shown in Figure \ref{mag-ori}. Since polarization angle (orientation of electric field) of synchrotron emission $\psi$ is perpendicular to orientation of magnetic field $\psi'$, Figure \ref{mag-ori} shows $\psi'$ maps, which were obtained by rotating the polarization angle by $90^\circ$ as $\psi' = \psi + 90^\circ$. Parameters of obtained maps are listed in Table \ref{tb:obs}, where Name, observation date, frequency, array configuration, size and position angle of synthesized-beam, and RMS (root-mean-square) noise levels in Stokes I, Q, and U maps can be referred. 
Two maps in two different bands are shown for the individual galaxies and it can be seen that the orientations of $\psi'$ change between two bands because of the Faraday rotation. Maps of higher frequency are closer to the intrinsic magnetic fields since the Faraday rotation is the smaller for higher frequency. Polarization angles were calculated only for areas where Stokes I has Signal-to-Noise (S/N) of more than $3\sigma$ and polarization intensity ($P$) has S/N of more than $3\sigma$. The RMS noise level in $P$ is also listed in Table \ref{tb:obs}. 

\begin{table*}[h]
  \caption{Basic Parameters of Samples}
  \label{tb:gal}
  \begin{center}
  \begin{tabular}{cccccccc}
    \hline\hline
    Name   & R.A.                                     & Dec.                     & Morph.     & B     & D           &inc.      &PA         \\
           &                                          &                          &            &       &[Mpc]&[$^\circ$]&[$^\circ$] \\ 
    (1)    & (2)                                      & (3)                      & (4)        & (5)   & (6)         & (7)      & (8)       \\
    \hline
    NGC4414& $12^{\rm h} 26^{\rm m} 27.5^{\rm s}$ & $+31^\circ 13^{'} 29^{''}$ & SA(rs)c    & 10.81 & 9.7        &50         &154        \\   
    NGC4736& $12^{\rm h} 50^{\rm m} 53.6^{\rm s}$ & $+41^\circ 07^{'} 10^{''}$ & (R)SA(r)ab & 8.80  & 4.3        &33         &119        \\   
    NGC6946& $20^{\rm h} 34^{\rm m} 52.0^{\rm s}$ & $+60^\circ 09^{'} 15^{''}$ & SAB(rs)bc) & 7.92  & 5.5        &42         &242        \\   
    \hline\hline
  \end{tabular}\\
(1) Name of galaxy, (2) (3) Right Ascension and Declination in J2000 \citep{dev91}, (4) Morphology \citep{dev91}, (5) Blue apparent magnitude \citep{tul88}, (6)  (7) inclination of disk \citep{tul88}, (8) Position angle of disk, which is defined as the angle between the orientations of major axis of disk and the north \citep{kun07}. 
  \end{center}
\end{table*}

\begin{table*}[h]
  \caption{Information of the VLA archive data}
  \label{tb:obs}
  \begin{center}
  \begin{tabular}{cccccccccc}
    \hline\hline
    Name    & Date       & Frequency & Array & Beam size     & BPA        & $\Delta$I & $\Delta$Q &$\Delta$U&$\Delta$P \\
            &            & [GHz]     &       & [$(^{''})^2$] & [$^\circ$] &\multicolumn{4}{c}{[mJy/beam]}    \\
    (1)     & (2)        & (3)       & (4)   & (5)           & (6)        & (7)       & (8)       & (9)      & (10)\\
\hline
    NGC4414 & 1995/06/03 &4.8851     &D      &$13\times 12$  &84          &0.095      &  0.014    &  0.013   &0.009\\
\cline{2-10}
            & 1995/06/04 &8.4149     &D      &$10\times 9 $  &85          &0.011      &  0.012    &  0.010   &0.007\\
\hline
    NGC4736 & 2007/04/17 &4.8851     &D      &$23\times 16$  &$-87$       &0.016      &  0.016    &  0.008   &0.008\\
\cline{2-10}
            & 2007/04/17 &8.4351     &D      &$13\times 10$  &$ 65$       &0.018      &  0.019    &  0.015   &0.011\\
\hline 
    NGC6946 & 1991/03/23 &4.8851     &D      &$18\times 18$  &$-69$       &0.039      &  0.012    &  0.012   &0.001\\
            & 1991/04/01 &           &       &               &            &           &           &          &     \\
\cline{2-10}
            & 1995/04/13 &8.4149     &D      &$28\times 26$  &$ 89$       &0.063      &  0.002    &  0.017   &0.015\\
            & 1995/04/20 &           &       &               &            &           &           &          &     \\
\hline
  \end{tabular}\\
(1) Name of galaxy, (2) Date of observation, (3) Observation frequency, (4) Array, (5) Synthesized beam size (Major axis $\times$  Minor axis), (6) Position angle of synthesized beam, (7) (8) (9) RMS noise levels in Stokes I, Q, and U, respectively, (10) RMS noise in Polarization intensity $P$. 
  \end{center}
\end{table*}

\section{Method for Making Magnetic Vector Maps}
Since the maps shown in Figure \ref{mag-ori} present orientations of the magnetic fields but not directions, magnetic vectors have the $180^\circ$-ambiguity. 
Here, we propose a new simple method for determining direction of magnetic field vector at each observed point using (1) geometrical orientation of disk (2) magnetic field orientation, and (3) sign of RM, as follows. 

\begin{figure}[h]
\centering
\rotatebox{0}{\includegraphics[width=7 cm]{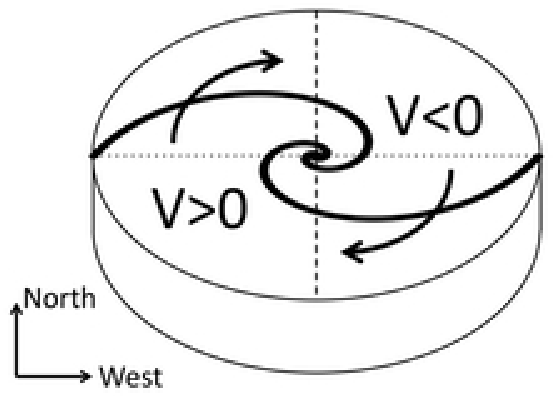}}
\rotatebox{0}{\includegraphics[width=7 cm]{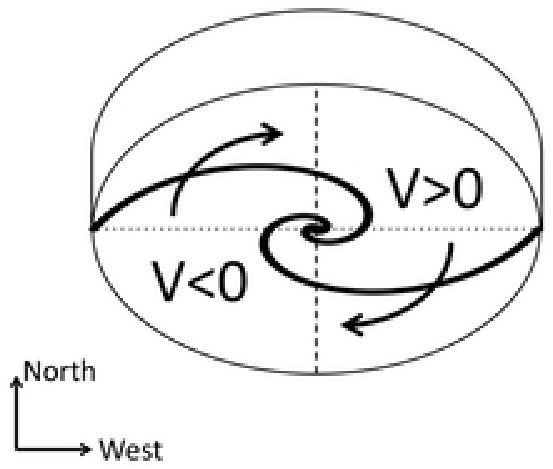}}
\caption{Determination of nearside of disk based on assumption of trailing spiral arms. }
\label{schematics1}
\end{figure}

\noindent{\bf (1) Geometrical Orientation of Disk}\\
Spiral galaxies generally have trailing spiral arms but not leading ones \citep{bin87}. If an image of spiral structure and velocity field are available, we can determine the geometrical orientation of a disk on an assumption of trailing spiral arms as follows. Figure \ref{schematics1} shows schematic illustrations of two galaxies with the same inclination, position angle (PA), and apparently the same spiral structure. For the simplicity, let us set the position angle to be 90$^{\circ}$. These model galaxies rotate in the clockwise as far as they have trailing spiral arms. 
If radial velocity of the western side is negative, the southern side is near side to the observer as shown in the left panel of Figure \ref{schematics1} because the model galaxy rotates in the clockwise. Similarly, if the radial velocity of the western side is positive, the northern side is near side to the observer as shown in the right panel of Figure \ref{schematics1}. 

Geometrical orientation of NGC4414, one of the selected galaxies in this paper, is that the north-eastern side is the near side since this galaxy rotates counter-clockwise assuming trailing spiral arms and the southern-eastern side is red-shifted consulting its velocity field \citep{kun07}. Similarly, the near side of NGC 4736's disk is the north-eastern side and that of NGC 6946 is the north-western side.

\begin{figure}[h]
\centering
\rotatebox{0}{\includegraphics[width=7 cm]{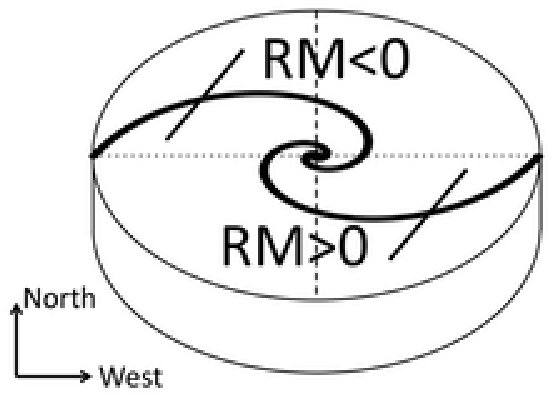}}
\rotatebox{0}{\includegraphics[width=7 cm]{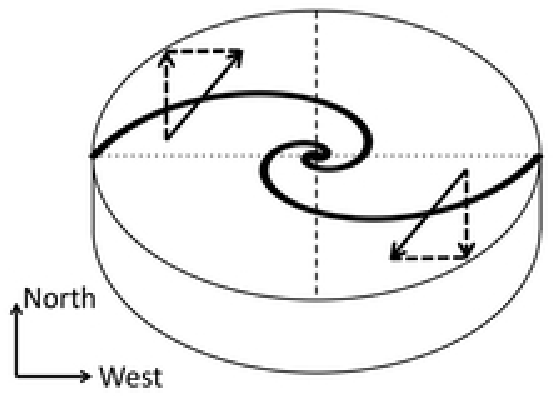}}
\caption{Determination of direction of magnetic field vector based on sign of the RM. }
\label{schematics2}
\end{figure}

\noindent{\bf (2) Transverse Magnetic Field Orientation}\\
The polarization angle less changed at higher frequency since the Faraday rotation is smaller for higher frequency. Therefore, maps of $\psi'$ at higher frequency can be taken for transverse magnetic field maps with the 180$^{\circ}$-ambiguity of in magnetic vector direction. The schematic illustration is shown in the left panel of Figure \ref{schematics2}. 
Ideally, maps of $\psi'$ can be replaced with the intrinsic magnetic field by subtracting the Faraday rotation using the RM. However, we adopt $\psi'$-map of higher frequency as a transverse magnetic field map for the simplicity. 

\noindent{\bf (3) Determination of the Direction of the Magnetic Field Vector}\\
Once the geometrical orientation of disk is determined as explained above, the 180$^\circ$-ambiguity can be resolved by seeing the line-of-sight directions of magnetic field, which can be derived with the sign of RM. 
Let us think a case of the left panel in Figure \ref{schematics1}. Left panel of Figure \ref{schematics2} shows that orientations of the magnetic field are observationally obtained at two points. When the RM is negative at the point in the north-eastern side (upper-left point), line-of-sight component of the magnetic field vector should be directed away from the observer (or directed northwards) as shown with a dashed arrow in the right panel of Figure \ref{schematics2}. Therefore, the magnetic vector has to be directed north-westwards (clockwise) but not south-eastwards (counter-clockwise). Thus, the 180$^{\circ}$-ambiguity can be resolved. 
In the same way, the line-of-sight component of magnetic field at the point in the south-western side (lower-right point) is directed toward us and the magnetic vector is directed south-eastwards (clockwise).
Thus, a magnetic field vector for each line of sight through the galaxy can be obtained. However, note that this does not work if a magnetic field vector has small RM. 

\section{Result}
\begin{figure}[h]
\centering
\rotatebox{0}{\includegraphics[height=5.5 cm]{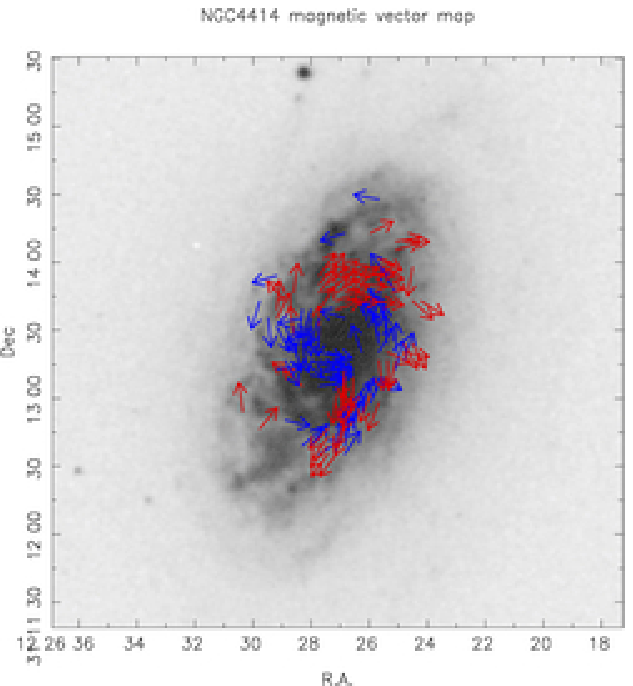}}
\rotatebox{0}{\includegraphics[height=5.5 cm]{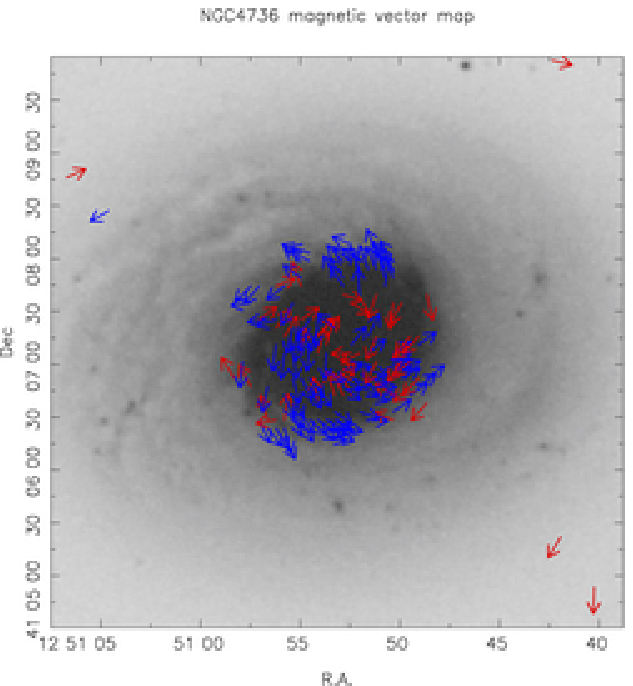}}
\rotatebox{0}{\includegraphics[height=5.5 cm]{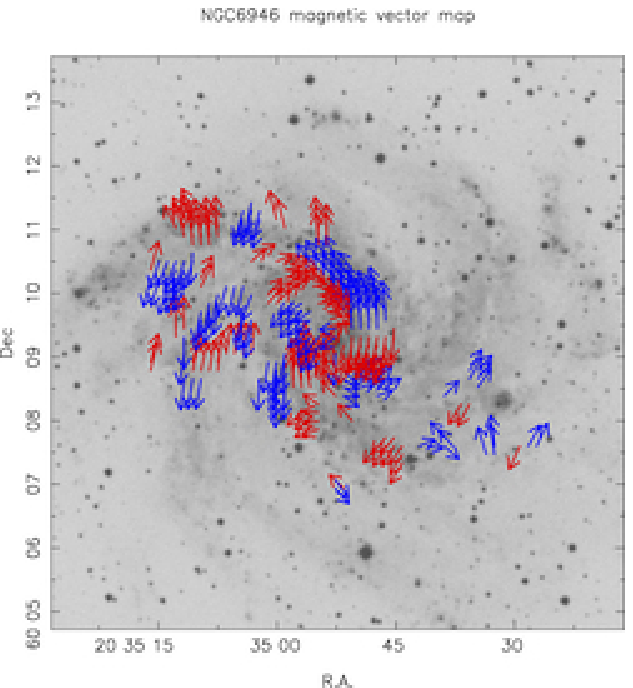}}
\caption{Magnetic field vector maps of NGC4414 (left), NGC4736 (middle), and NGC6946 (right) obtained by applying our new method. Background gray scale images are taken from DSS. Red and blue vectors denote clockwise and counter-clockwise vectors, respectively. }
\label{mag-vec}
\end{figure}

Obtained vector maps for the three galaxies are shown in Figure \ref{mag-vec}. Vectors are plotted only at points where Stokes I is larger than $10\sigma$, polarization intensity is larger than $3\sigma$, and error of orientation is smaller than angle of the Faraday rotation. Therefore, the number of vectors is smaller than the original polarization maps. Red and blue vectors denote clockwise and counter-clockwise vectors, respectively. 

We also show face-on views of magnetic field vector maps for these three galaxies in Figure \ref{faceon-mag-vec}. Circles are superposed on the maps to see radial variations of vectors.

\section{Discussion}
   
\subsection{Merits and limits of our Proposed Method}
Merit of our proposed method is that magnetic vector for each line of sight through the galaxy can be obtained in an inexpensive way, without any assumptions on configurations of magnetic fields but only with simple a reasonable assumption of trailing spirals and an assumption that Faraday rotation is dominant at disk. 
Optical images and velocity fields are now easily available to determine geometrical orientation of disk. And polarization data at two different bands are also easily available from archival systems such as the VLA. Therefore, it is expected that our method is useful and has a potential to promote study on configuration of magnetic fields. 

However, note that this method should be used with a careful consideration about limits. The first limit is that this method should not be used in the case that RM is small. If the RM is comparable to the RMS noise, direction of vector is affected by noise and cannot be determined. Therefore, we took only points where errors of orientations are smaller than Faraday rotation angle. 
Because of this effect, magnetic vectors with small RM need to be omitted.  

The second limit is that face-on galaxies is not suitable for applying this method because vertical component of magnetic field in halo affect the Faraday rotation. However, such an effect of vertical magnetic field are negligible in the case that inclination angle is larger than $15^\circ$ according to \citet{ste08}. The latter effect seems negligible for the three selected galaxies in this paper since the inclination range is $30^\circ$--$60^\circ$. 

The third limit is that this method is suitable for studies on the mean magnetic fields in galaxies but not for studies on the total magnetic field. For the latter case of the total magnetic field, total intensity (Stokes I) should be also taken into account.

\subsection{Configurations of magnetic fields in three galaxies}
Figures \ref{mag-vec} and \ref{faceon-mag-vec} show that all three galaxies include both clockwise and counter-clockwise vectors within them and that there is no galaxy showing simple ASS mode among the samples. Therefore, magnetic reversals seem to commonly exist and higher modes such as BSS or QSS are not negligible in general. It should be noted that the data shown in this paper are taken from only the interferometry and they generally suffer the missing flux due to lack of zero-spacing baseline. Since all the data were obtained in D-array whose shortest baseline was 35m, emission from larger structure than $0.06 {\rm (m)}/35 {\rm (m)}\sim 5.9'$ was resolved out in C-band and $0.036 {\rm (m)}/35 {\rm (m)}\sim 3.5'$ in X-band. We assumed the same rate of the missing flux of the Stokes Q and U for the simplicity because we focus on proposing the method in this paper and most of the emissions are concentrated within radius of 3.5' from centers as shown in Figure \ref{mag-ori}. Therefore, it should be noted that larger-scale component than $3.5'$ were not traced in this paper even if it exists. 
Details for individual galaxies are described below.  

\subsubsection{NGC 4414}
NGC 4414 is known as an isolated flocculent spiral galaxy without strong density wave flows. Despite of such a weak density wave, a coherent magnetic field is clearly observed and its pitch angle is almost the same as that of the optical spiral structures \citep{soi02}. \citet{soi02} studied the azimuthal variations of magnetic pitch angles and concluded that it can be explained with mixture of ASS, BSS, and QSS, whose relative strengths are 1:0.6:0.3--0.5. 
 
In our study, the obtained magnetic vector map of NGC4414 (left panel of Figure \ref{faceon-mag-vec}) shows that clockwise and counter-clockwise magnetic vectors exist. Therefore, we can conclude that the magnetic field has no Axis-Symmetric components as the former work suggested. Left panel of Figure \ref{faceon-mag-vec} shows that an annulus between the second and third circles from the center has clockwise vectors in the top-left and bottom-right regions, and has counter-clockwise ones in the rest. The mode of QSS seems dominant in the global magnetic field since there are four reversals in a single annulus and other annuli have the similar patterns. However, relative strength of the magnetic field configurations is not concluded in this paper since more studies including careful comparisons with theoretical models should be carried out and are not focuses of this paper.

\subsubsection{NGC 4736}
NGC 4736 (M94) is known as a nearby isolated spiral galaxy with two ring-like structures. Synchrotron polarization observations showed that this galaxy has an ordered magnetic field with the spiral pattern, which is not coupled with its gaseous distribution or motion \citep{chy08}. The RM is large in the northern region of the disk and this asymmetry is thought to indicate existence of action of large-scale magneto-hydrodynamic (MHD) dynamo. 
 
Looking at the magnetic vector map of this galaxy (middle panel of Figure \ref{faceon-mag-vec}), both clockwise and counter-clockwise magnetic vectors can be found. This fact implies that higher modes such as BSS ($m=1$) should exist in the magnetic field. An annulus between the first and second circles has clockwise vectors in top-right, bottom, and top-left parts and counter-clockwise ones in the others. Naively accepting this six-times changes of magnetic reversals, the Sexi-Symmetric Spiral (SSS) mode of $m=3$ might be dominant in this region.

\subsubsection{NGC 6946}
NGC 6946 is a nearby grand design spiral galaxy with symmetric magnetic spiral arms between the optical spiral arms \citep{bec96}. \citet{bec07} studied the magnetic field of this galaxy in detail and concluded that the inner two magnetic arms are superposition of dynamo modes of $m=0$ and $m=2$ and that the the inner and outer magnetic fields are amplified by turbulent gas flow and Magneto-rotational instability, respectively.  

\begin{figure}[h]
\centering
\rotatebox{0}{\includegraphics[height=5.5 cm]{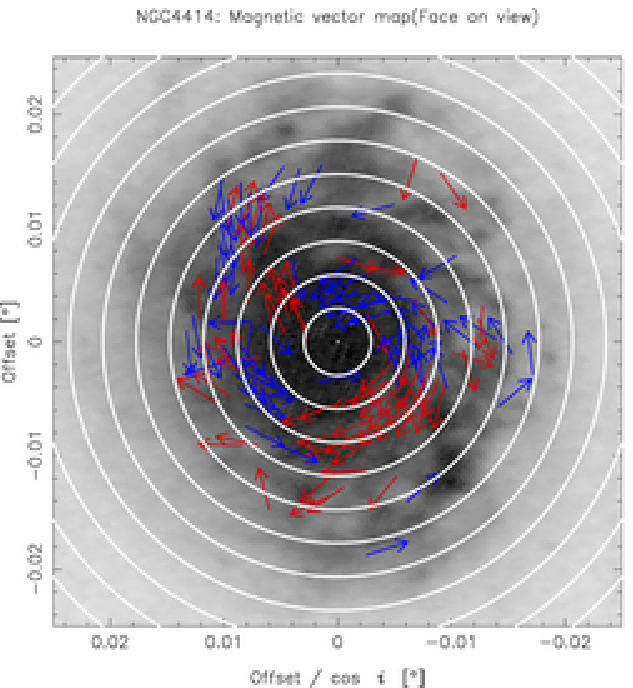}}
\rotatebox{0}{\includegraphics[height=5.5 cm]{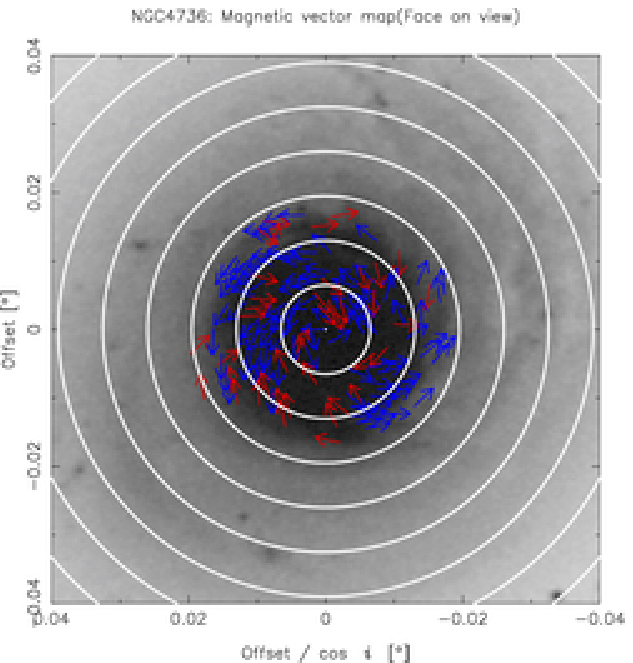}}
\rotatebox{0}{\includegraphics[height=5.5 cm]{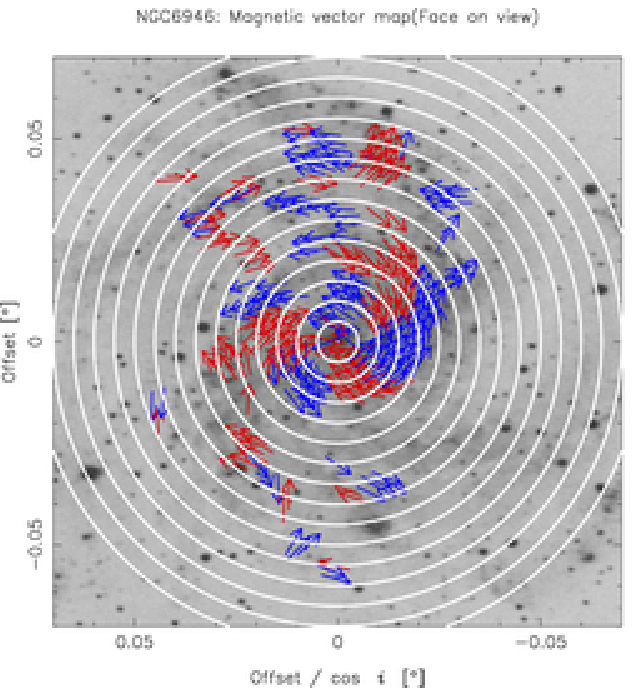}}
\caption{Face-on views of magnetic field vector maps of NGC 4414 (left), NGC 4736 (middle), and NGC 6946 (right), superposed on the optical B-band images taken from DSS. Each image was rotated by Position angle and horizontally enlarged by $1/\cos{i}$. Circles are overlaid at 500-pc intervals. }
\label{faceon-mag-vec}
\end{figure}

Our study shows that both clockwise and counter-clockwise magnetic vectors exist to support existence of higher dynamo modes as shown in Figures \ref{mag-vec} and \ref{faceon-mag-vec}. Right panel of Figure \ref{faceon-mag-vec} shows that annuli between the first and fifth circles contain six reversals in the azimuthal direction. This fact seems to imply that this galaxy also has the SSS magnetic field. According to \citet{kur18}, the reversals occur four-times in inter-arm regions and twice at spiral arms. 

\section{Summary}
We present a new method for deriving magnetic vector field maps for spiral galaxies based on simple assumptions and relatively easily available data. All data we need are two synchrotron polarization maps obtained in two different bands, an optical image, and velocity field, which is obtained by line-observations such as HI, CO, or H$\alpha$. First, geometrical orientations of disks are determined by consulting an optical image and velocity field based on the assumptions that spiral pattern is generally trailing and that Faraday rotation is dominant in disk. Orientations of magnetic field lines are given by polarization map of the higher band and the 180$^\circ$-ambiguity of magnetic field vector is solved by checking sign of the RM. Thus, a magnetic field vector for each line of sight through the galaxy can be derived.

We selected three galaxies, NGC 4414, NGC 4736, and NGC 6946, to apply this method and carried out polarization data reduction for archival data released from the VLA. We found that all three galaxies have both clockwise and counter-clockwise vectors, that no galaxies out of three has a mode of simple ASS magnetic field, and that higher modes such as BSS, QSS, and SSS seem to exist. 

This method should be used with cares that face-on and edge-on galaxies should be avoided from applying. And the magnetic vectors with small RM need to be omitted.  However, our proposed method has a merit to derive a magnetic field vector for each line of sight through the galaxy.

\vspace{6pt} 
\authorcontributions{Conceptualization, Methodology, and Investigation, H.N. and K.A.; Validation, H.N., K.K. and K.A.; Writing – Original Draft Preparation, H.N.; Visualization, H.N. and K.K.}
\funding{This research received no external funding.}
\acknowledgments{We thank the anonymous referees for their invaluable comments, which have greatly improved this paper. We also thank participants to the international workshop entitled with ``THE POWER OF FARADAY TOMOGRAPHY --- TOWARDS 3D MAPPING OF COSMIC MAGNETIC FIELDS ---'', for giving useful comments and discussion to our study. }
\conflictsofinterest{The authors declare no conflict of interest.} 
\reftitle{References}

\end{document}